\documentclass[journal]{IEEEtran}
\usepackage[utf8]{inputenc}
\UseRawInputEncoding

\usepackage{comment}
\usepackage{hyperref}
\hypersetup{ 
 colorlinks=true, 
 linkcolor=red, 
 filecolor=blue, 
 citecolor = blue, 
 urlcolor=blue, 
} 

\usepackage{graphicx}
\usepackage[utf8]{inputenc}

\usepackage{subcaption}
\usepackage[acronym, nonumberlist, nohypertypes={acronym}]{glossaries}
\newacronym{uav}{UAV}{Unmanned Aerial Vehicle}
\newacronym{vr}{VR}{Virtual Reality}
\newacronym{ar}{AR}{Augmented Reality}
\newacronym{llm}{LLM}{Large Language Model}
\newacronym{ai}{AI}{Artificial Intelligence}
\newacronym{ml}{ML}{Machine Learning}
\newacronym{urllc}{URLLC}{Ultra-Reliable Low-Latency Communication}
\newacronym{embb}{eMBB}{Enhanced Mobile Broadband}
\newacronym{mmtc}{mMTC}{Massive Machine-Type Communications}
\newacronym{ntn}{NTN}{Non-Terrestrial Network}

\begin{document}
%
\title{Beyond Visual Line of Sight:  UAVs with Edge AI, Connected LLMs, and VR for Autonomous Aerial Intelligence}
%
%
%


\author{
\IEEEauthorblockN{
Andr\'{e}s~Navarro\IEEEauthorrefmark{1}, 
Carlos~de~Quinto\IEEEauthorrefmark{1}, 
Jos\'{e}~Alberto~Hern\'{a}ndez\IEEEauthorrefmark{1} 
}
\\
\IEEEauthorblockA{\IEEEauthorrefmark{1} Dept. Telematics Engineering, Universidad Carlos III de Madrid, Spain}\\
}

\maketitle

\begin{abstract}
Unmanned aerial vehicles (\glspl{uav}) are reshaping non-terrestrial networks (\glspl{ntn}) by acting as agile, intelligent nodes capable of advanced analytics and instantaneous situational awareness. This article introduces a budget-friendly quadcopter platform that unites 5G communications, edge-based processing, and artificial intelligence (\gls{ai}) to tackle core challenges in \gls{ntn} scenarios. Outfitted with a panoramic camera, robust onboard computation, and large language models (\glspl{llm}), the drone system delivers seamless object recognition, contextual analysis, and immersive operator experiences through virtual reality (\gls{vr}) technology. Field evaluations confirm the platform’s ability to process visual streams with low latency and sustain robust 5G links. Adding LLMs further streamlines operations by extracting actionable insights and refining collected data for decision support. Demonstrated use cases, including emergency response, infrastructure assessment, and environmental surveillance, underscore the system’s adaptability in demanding contexts. A demonstration video is available at \url{https://youtu.be/F1fDpXw-kBg}, showcasing real-world operation.
\end{abstract}


%
\IEEEpeerreviewmaketitle

\section{Introduction}
The development of intelligent unmanned aerial vehicle (UAV) platforms that seamlessly integrate multiple advanced technologies presents significant engineering challenges that require systematic methodological approaches. While individual technologies such as 5G communications, edge computing, and virtual reality have matured independently, their convergence in resource-constrained aerial platforms demands careful consideration of integration strategies. The complexity of combining heterogeneous systems—each with distinct power requirements, processing demands, and communication protocols—necessitates a structured development framework that addresses both technical compatibility and operational reliability\cite{borcoci2024uav}.

Current UAV development approaches often focus on single-technology implementations or present end-to-end solutions without detailing the underlying methodological foundations that enable successful integration. This gap limits the reproducibility of research outcomes and hinders the advancement of multi-technology UAV platforms in both academic and industrial contexts. The engineering challenges span multiple domains: hardware integration must balance computational performance with weight and power constraints, software architectures must coordinate real-time processing, and communication protocols must maintain reliability under dynamic network conditions.

The methodology presented in this work addresses these challenges for developing UAV platforms that integrate 5G connectivity, edge-based artificial intelligence, and immersive virtual reality. This approach encompasses component selection criteria, modular integration strategies, and validation procedures that ensure reliable operation across all subsystems.

\begin{figure}
    \centering
    \includegraphics[width=1\linewidth]{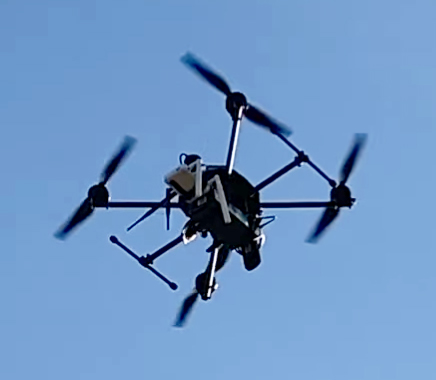}
    \caption{Prototype UAV platform demonstrating the integrated multi-technology architecture developed using the proposed methodology.}
    \label{fig:Flying_Drone}
\end{figure}

This work demonstrates the successful convergence of cutting-edge technologies—5G communications, edge AI, and immersive VR—within a single UAV platform, showcasing the practical feasibility of multi-technology integration in aerial systems. The implementation reveals both the potential and the challenges, providing valuable insights into hardware optimization, software architecture design, and real-world performance trade-offs. Through detailed technical implementation and experimental validation, this platform serves as a proof-of-concept that bridges the gap between individual technology capabilities and their unified application in intelligent aerial systems.

\section{Related Work and Comparative Analysis}\label{sec:related_work}

Recent UAV research has explored individual technology integrations, with limited focus on comprehensive multi-technology platforms. Current UAV-AI implementations primarily address natural language command processing and mission planning~\cite{javaid2024large}, while edge computing approaches concentrate on computational offloading strategies~\cite{janßen2023supportinguavsedgecomputing}. VR-UAV integration efforts focus predominantly on teleoperation and first-person control interfaces~\cite{freeman2025vr}, lacking intelligent analytics capabilities.

Our platform differs significantly by providing unified integration of edge AI processing, immersive VR interfaces, and intelligent analytics within a single cost-effective system. Unlike existing solutions that implement individual technologies, our approach combines real-time YOLOv11 object detection, contextual LLM analysis, and 360° VR streaming in a distributed architecture. The system addresses practical implementation challenges, including hardware constraints, thermal management, and network optimization that are often overlooked in theoretical frameworks. Furthermore, our comprehensive web application with NextJS framework, database management, and automated alert generation represents a complete operational solution rather than a proof-of-concept demonstration.

\section{System Architecture and Technical Implementation} \label{sec:Design}

The development of an intelligent UAV platform that seamlessly integrates 5G communications, edge computing, virtual reality technologies, and large language models (\glspl{llm}) requires careful consideration of numerous technical trade-offs and architectural decisions. This section details the rationale behind key design choices, component selection criteria, and integration strategies that enable the successful convergence of heterogeneous technologies within the constraints of aerial deployment. A more detailed explanation can be found in previous research ~\cite{caceres2024developingcosteffectivedrones5g} 

The platform architecture addresses three fundamental challenges: maintaining computational performance while minimizing weight and power consumption, and achieving real-time processing capabilities across distributed systems. Each design decision was evaluated against multiple criteria, including cost-effectiveness, technical compatibility, operational reliability, and scalability potential.

The following subsections examine the hardware architecture decisions, multi-technology integration strategies, distributed processing design, and communication infrastructure choices that collectively enable the platform's advanced capabilities. Rather than presenting a prescriptive methodology, this analysis focuses on the engineering rationale and performance trade-offs that shaped the final system configuration, providing insights into the practical considerations that govern multi-technology UAV development.


\subsection{Hardware Architecture and Component Selection}

The hardware architecture design prioritized three critical factors: computational performance for real-time AI processing, communication capabilities for 5G/4G connectivity, and cost-effectiveness to demonstrate the feasibility of intelligent UAV platforms without prohibitive investment. Each component selection involved careful trade-off analysis between performance requirements, weight constraints, and power consumption limitations.

The NVIDIA Jetson Orin Nano was selected as the primary computing platform due to its exceptional AI performance-to-power ratio, delivering 40 TOPS of AI compute while consuming only 15W. This choice addressed the fundamental challenge of running YOLOv11 inference at acceptable frame rates while maintaining flight endurance. Alternative solutions, such as Intel NUC or Raspberry Pi 4, were evaluated but rejected due to insufficient GPU acceleration capabilities for real-time computer vision tasks.

The Ricoh Theta X 360° camera selection balanced image quality requirements with integration complexity. Its 4K resolution capability and direct USB streaming interface eliminated the need for complex video capture hardware. The decision to use 360° imaging over traditional fixed-angle cameras addressed the operational requirement for \gls{vr} situational awareness without mechanical gimbal systems that would increase weight and complexity.

Despite the project's 5G focus, the GL-X750V2 4G router was selected due to local availability and cost considerations. This pragmatic decision demonstrated that advanced UAV capabilities remain achievable with current 4G infrastructure, while the router's dual-band Wi-Fi and Ethernet connectivity provided flexible integration options for the distributed processing architecture.



The physical integration process revealed several critical considerations. Thermal management did not become an issue when operating the Jetson Orin Nano, since drone-mounted components have high amounts of airflow due to the propeller rotation, flight height, and movement speed. Power distribution complexity increased significantly with the addition of multiple devices with different voltage necessities, having to accommodate several regulators. Weight distribution optimisation ensured that the additional payload did not adversely affect flight characteristics or centre of gravity positioning.

\subsection{Physical Integration and Mounting Strategy}

\begin{figure}
    \centering
    \begin{subfigure}[b]{0.8\columnwidth}
        \includegraphics[width=1\linewidth]{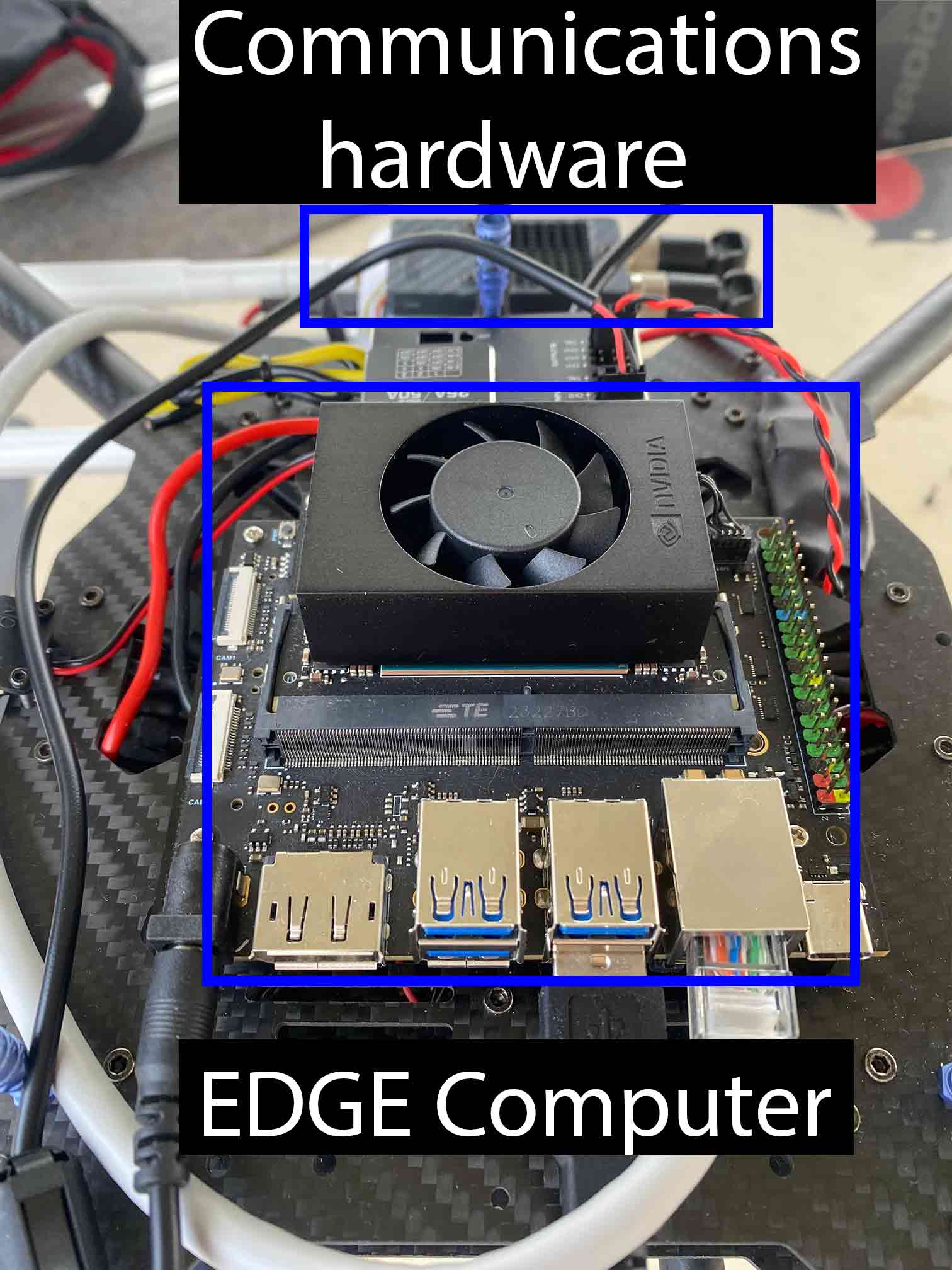}
        \caption{Edge computing and communications hardware placement optimized for thermal dissipation.}
        \label{fig:drone_edge_computer}
    \end{subfigure}
    \begin{subfigure}[b]{0.8\columnwidth}
        \includegraphics[width=1\linewidth]{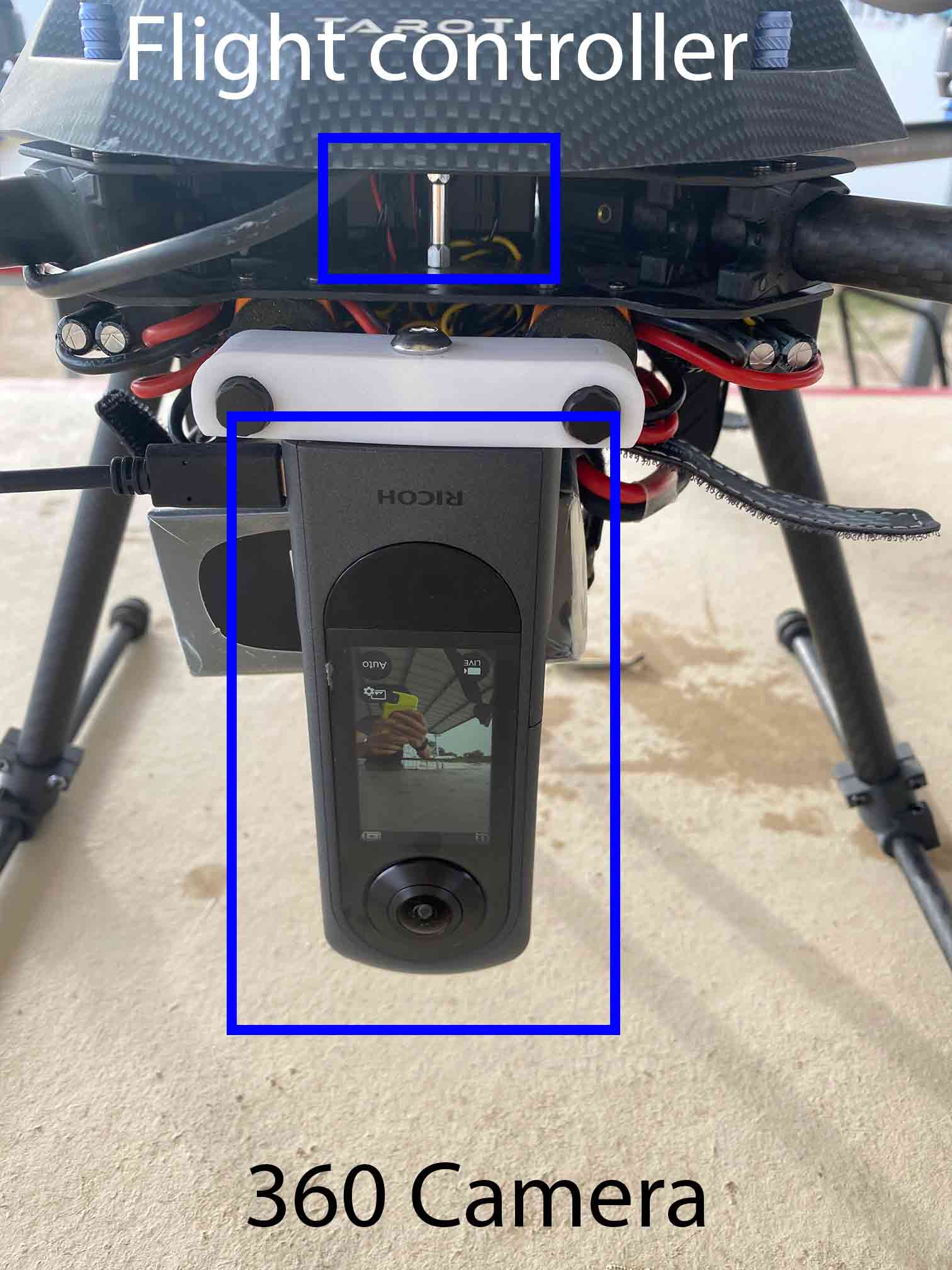}
        \caption{Camera positioning for unobstructed 360° coverage while maintaining center of gravity balance.}
        \label{fig:camera_installed}
    \end{subfigure}
    \caption{Integrated multi-technology payload demonstrating strategic component placement.}
    \label{fig:drone}
\end{figure}

The integration of multiple subsystems required careful consideration of weight distribution and physical space. Figures~\ref{fig:drone_edge_computer} and~\ref{fig:camera_installed} illustrate the final configuration that balances these competing requirements.

The NVIDIA Jetson Orin Nano's elevated mounting position maximizes airflow, accessibility, and protection. The Ricoh Theta X central placement ensures good 360° coverage. The GL-X750V2 router positioning provide the best place for antennas to obtain optimal signal reception.

\begin{figure}[h!]
    \centering
    \includegraphics[width=1\linewidth]{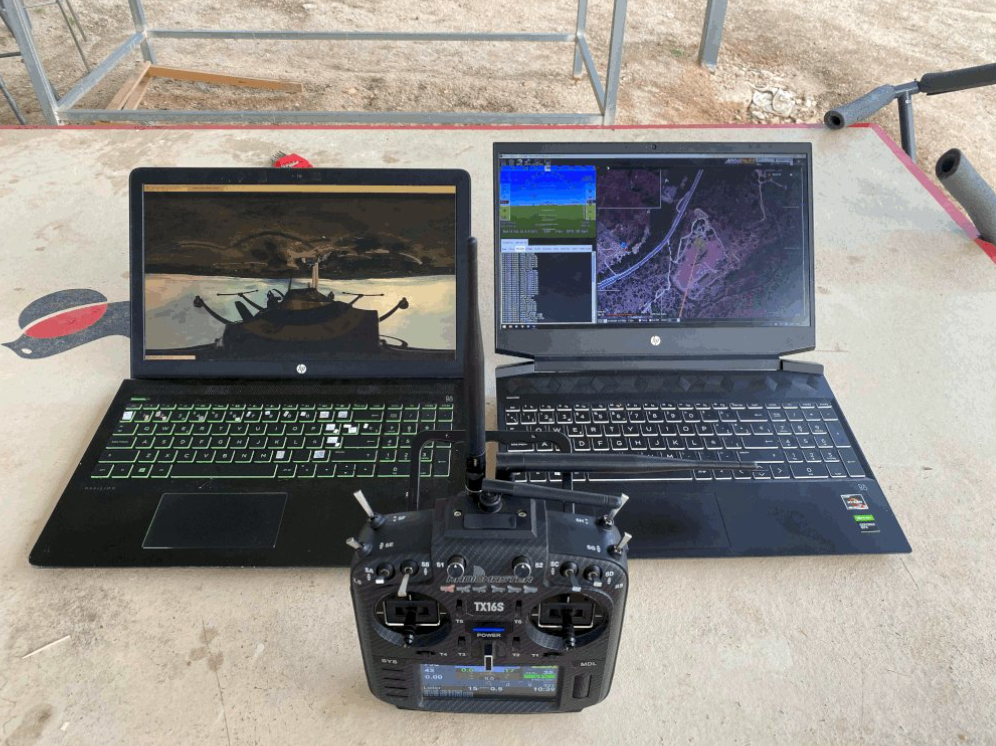}
    \caption{Multi-computer ground control station designed for redundant operation and pilot intervention capability.}
    \label{fig:control station}
\end{figure}

The drone is operated from a comprehensive control station featuring three essential components can be seen at figure \ref{fig:control station}. The primary computer handles flight programming, route setting, and telemetry monitoring, providing real-time data on the drone's status and position. The secondary computer manages the advanced AI and VR capabilities, processing detection alerts and controlling the virtual reality service for enhanced situational awareness. Completing the system is a dedicated controller unit housing the communication module, which not only transmits commands to the drone but also serves as a critical failsafe mechanism, allowing the pilot to immediately override autonomous functions and take direct manual control when necessary.

A complete system demonstration is available at \url{https://youtu.be/F1fDpXw-kBg}.

\subsection{System Architecture}

At the edge, the Jetson Orin Nano runs YOLOv11 for object detection and encodes 360° video using GStreamer, using cuda for 5 FPS inference.The middleware employs ZeroTier for a resilient virtual network, establishing secure overlay connectivity that operates independently of telecommunications carrier routing and enables direct peer-to-peer communication between system components.

Cloud processing uses Groq’s API for LLM inference. When the edge detects an object, images are compressed and sent to the cloud, where Llama-3.2-90b-vision-preview generates natural language analyses, returned promptly to the operator. This offloading enables advanced vision-language features without overloading the edge.

\begin{figure}[h!]
    \centering
    \includegraphics[width=1\linewidth]{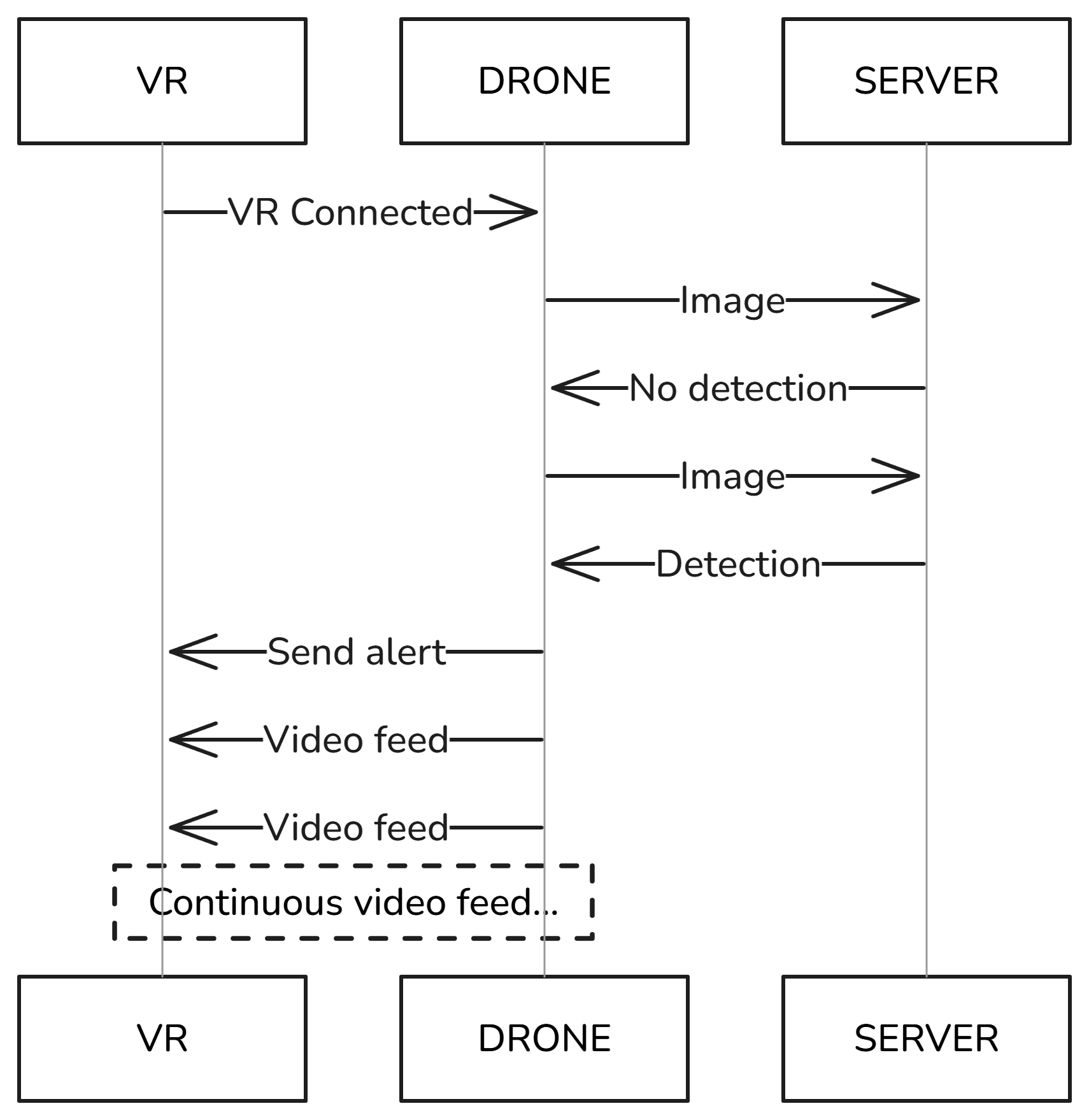}
    \caption{Sequence diagram: The drone sends images to the server. On detection, the drone notifies the VR system (left), which then receives a live video stream.}
    \label{fig:Detection sequence diagram}
\end{figure}

Finally, the VR Headset is connected to the drone hosting a WebSocket server where the live video feed can be accessed. To connect to this WebSocket server, a Unity application has been developed using the Oculus framework. An image of this application can be seen in Figure \ref{fig:AR Keyboard}. Figure~\ref{fig:Detection sequence diagram} illustrates the system workflow: the drone transmits an image to the server, which analyzes it for detections. If no detection is found, the server notifies the drone, and the process repeats. Upon a positive detection, the drone alerts the VR system (leftmost in the diagram), and live video is then broadcast to the VR headset for immersive monitoring.

This flow enables efficient conditional streaming, ensuring that high-bandwidth video is only transmitted to the VR interface when relevant detections occur, similar to selective end-to-end streaming architectures described in recent UAV surveillance literature~\cite{TowardshighVideostreaming, ARflightAssistance}.

\subsection{Distributed AI Pipeline and Web Application Integration}

The UAV reconnaissance platform employs a comprehensive web application built with the NextJS framework~\cite{nextjs2024}, managing the complete intelligence pipeline from object detection to contextual analysis. The system architecture consists of three distinct layers: a ReactJS presentation layer with Tailwind CSS styling for user interface management~\cite{reactjs2024}~\cite{tailwindcss2024}, an application layer handling business logic including mission control and rule-based alert generation, and a data layer utilizing Amazon S3 for object storage~\cite{aws_s3_2020} combined with a flexible database schema supporting PostgreSQL, MySQL, or SQLite through Prisma ORM~\cite{prisma2024}.

The server infrastructure comprises dual processing units that handle object detection verification and alert system management. When objects are detected by the UAV's onboard systems, images are transmitted securely through a VPN tunnel to the central server, which processes the data through advanced language model integration using Llama 3.2 90b Vision~\cite{meta2024llama}. The LLM component receives detection data with operator-specified prompts, generating a comprehensive analysis that includes object classification details, behavioral patterns, and potential security implications. This contextual analysis enhances operator decision-making by providing detailed descriptions, identifying relationships between detected objects, and generating natural language summaries of surveillance data.

\begin{figure*}[!ht]
    \centering
    \includegraphics[width=\textwidth]{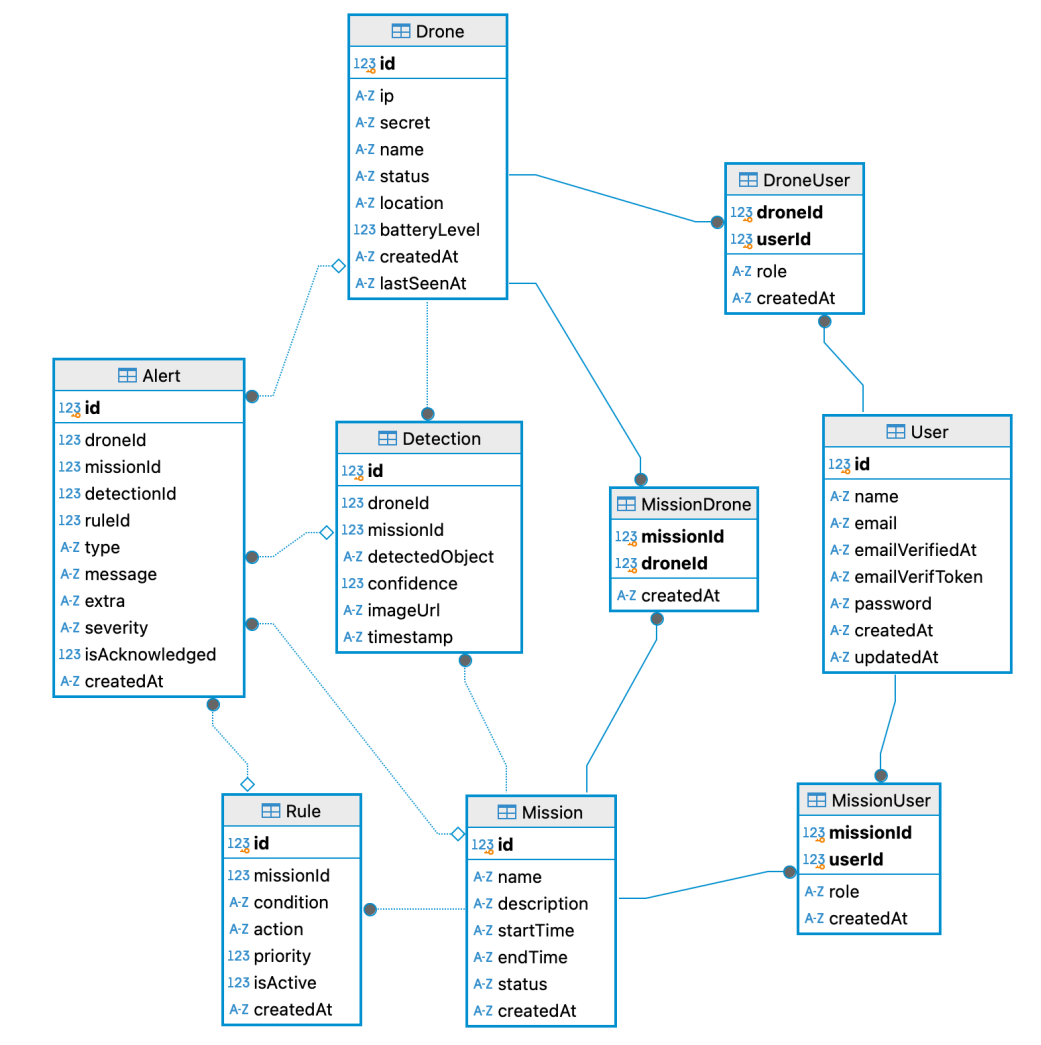}
    \caption{Database schema of the reconnaissance platform showing the relational structure for users, drones, missions, detections, and alerts.}
    \label{fig:dbdiagram}
\end{figure*}

The database schema, see Figure~\ref{fig:dbdiagram}, encompasses six main tables that collectively enable comprehensive UAV fleet management and automated alert generation. The user table manages authentication and authorization for system access using NextAuth.js~\cite{nextauth2024}, while the Drone table stores unique identifiers and secret tokens for each connected UAV. The Mission table organizes drone operations and groups UAVs for specific tasks, with users assigned to missions for data access control. The Rule table defines mission objectives and constraints, specifying detection targets and alert parameters. The Detection table stores comprehensive object information, including class, confidence, position, size, and orientation data. Finally, the Alert table manages system-generated notifications with properties such as type, severity, message, timestamp, and status, enabling real-time situational awareness and automated response protocols.

A complete system demonstration is available at \url{https://youtu.be/kdMgwRRte-8}.

\subsection{VR Streaming Implementation}
\begin{figure}
    \centering
    \includegraphics[width=1\linewidth]{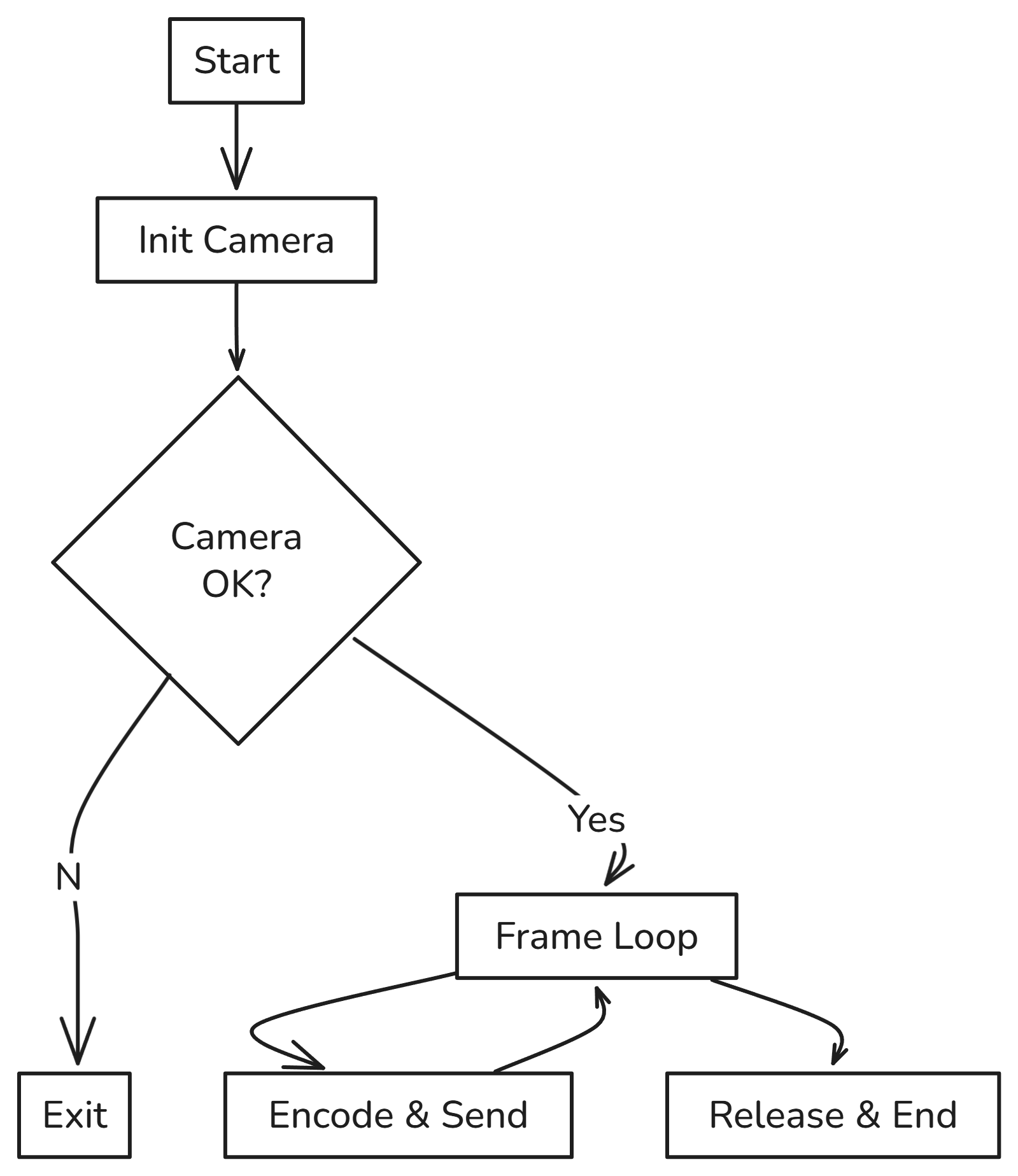}
    \caption{Live video streaming workflow: camera status verification, continuous frame processing, and conditional encoding/transmission via WebSocket protocol.}
    \label{fig:Code Flowchart}
\end{figure}

The immersive video system employs a three-stage pipeline for live drone monitoring. The Ricoh Theta X streams 360° video to an NVIDIA Jetson Orin Nano, which encodes the feed using H.264 compression. Encoded frames transmit through a custom Python WebSocket server as base64-encoded messages.  A flowchart for this code can be seen at ~\ref{fig:Code Flowchart}.

\begin{figure}
    \centering
    \includegraphics[width=1\linewidth]{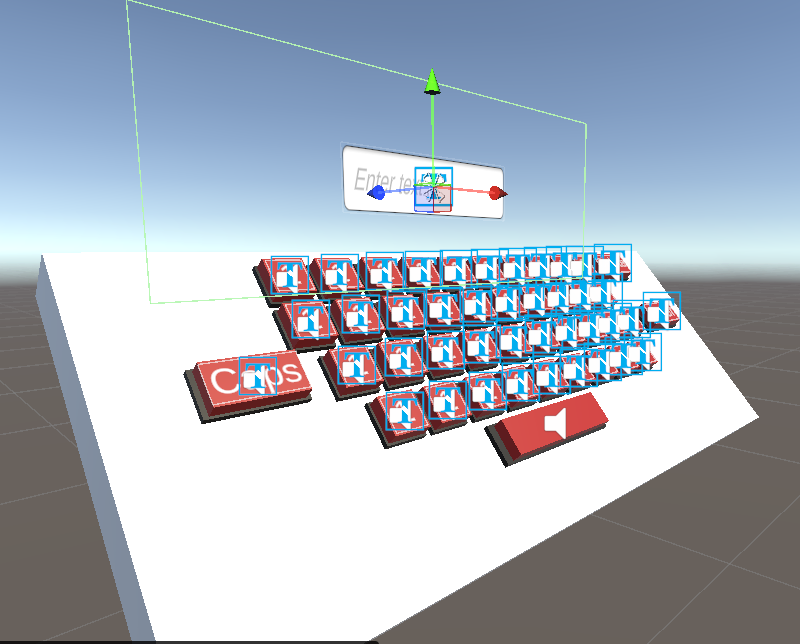}
    \caption{VR Environment: Mixed reality keyboard for settings control}
    \label{fig:AR Keyboard}
\end{figure}

On the client side, a Unity 6 application running on Meta Quest 3 headsets receives and decodes the video stream. The implementation leverages Meta's Oculus XR framework for hand tracking and video passthrough capabilities, enabling users to select drones using hand recognition in a mixed reality environment, as shown in Figure~\ref{fig:AR Keyboard}.

The system exhibits an average end-to-end latency of 10.2 seconds, which remains acceptable for non-critical monitoring applications. Operators can freely pan their viewing perspective within the 360° video stream by moving their head.

A field demonstration of the VR view is available at \url{https://youtube.com/shorts/eilMtAnVEM8}.

\begin{figure}
    \centering
    \begin{subfigure}[b]{0.8\columnwidth}
        \includegraphics[width=1\linewidth]{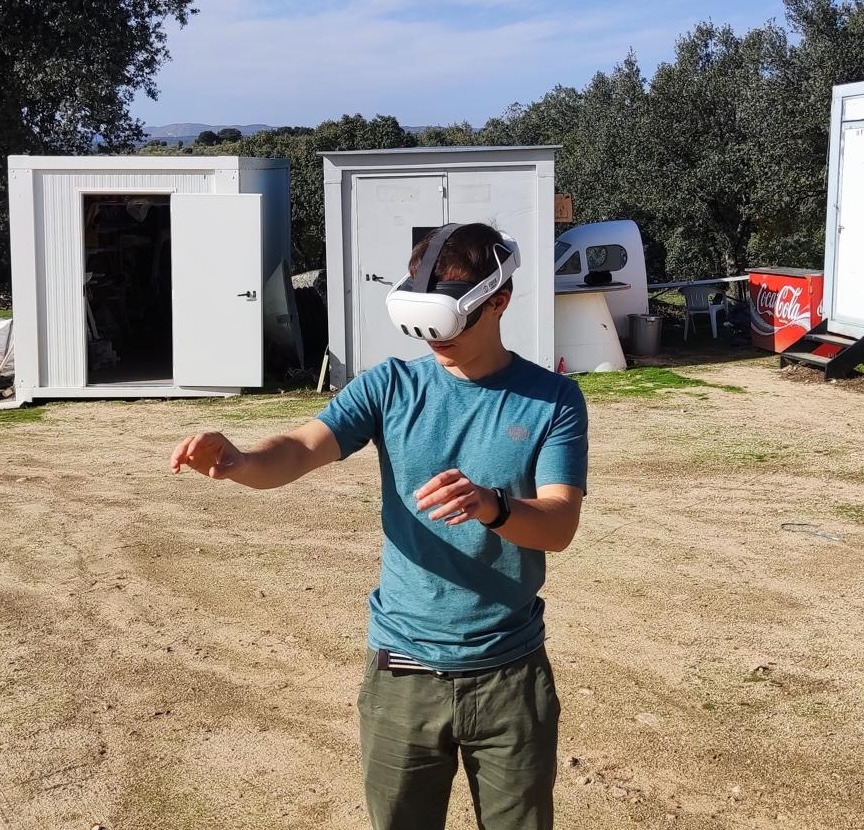}
        \caption{Operator using VR headset in field test.}
        \label{fig:vr_participant}
    \end{subfigure}
    \begin{subfigure}[b]{0.8\columnwidth}
        \includegraphics[width=1\linewidth]{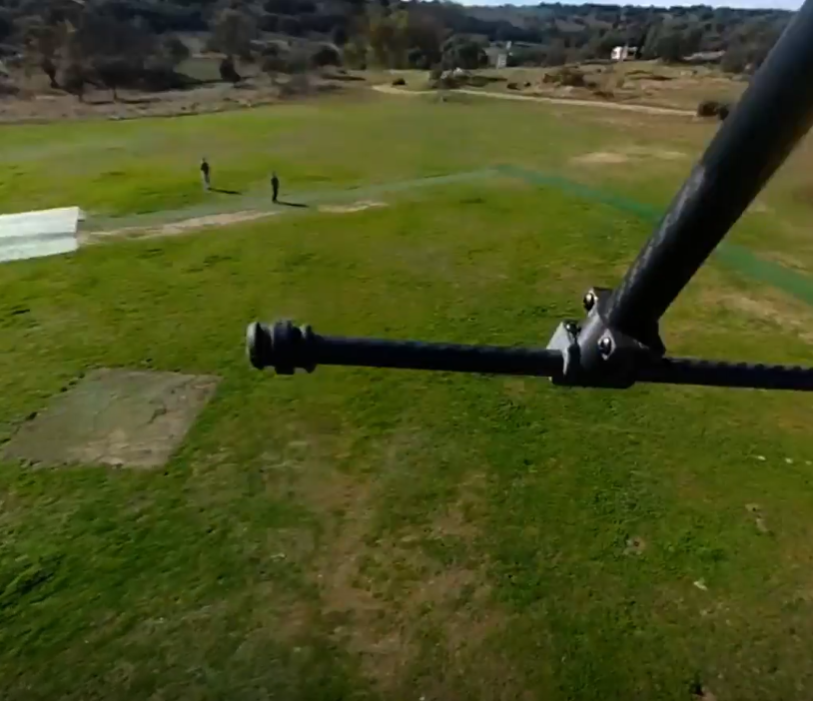}
        \caption{Live VR headset view of 360° stream.}
        \label{fig:vr view}
    \end{subfigure}
    \caption{VR streaming system in field evaluation.}
    \label{fig:vr_main}
\end{figure}

\section{Experimental Evaluation}\label{sec:experiments}

This section details the experimental validation of the UAV’s AI and VR features, focusing on two scenarios: (1) live human detection with LLM-based analysis, and (2) immersive VR monitoring using Meta Quest 3. All tests were conducted in real-world settings. Source code is available at \url{https://github.com/orgs/6G-Integration-3-UC3M/repositories}.

\subsection{AI Performance Assessment}

Field testing revealed comprehensive insights into the integrated AI pipeline's operational capabilities and limitations. The YOLOv11 implementation achieved 0.72 average precision (AP@0.5) for human detection with per-frame processing times of 0.2–0.4 seconds on the NVIDIA Jetson Orin Nano, enabling sustained 5 FPS inference rates under full computational load. The combined AI workflow, incorporating both edge detection and cloud-based LLM analysis, achieved total response times averaging 2.6 seconds from initial detection to contextual analysis delivery, consisting of YOLOv11 inference (0.3s), image compression and transmission (0.8s), Llama-3.2-90b-vision-preview processing (2.1s), and response delivery (0.4s). Analysis revealed several bottlenecks limiting overall system performance: network transmission delays accounted for 30\% of total response time, suggesting opportunities for edge-cloud load balancing optimization, while the current H.264 encoding pipeline introduced compression artifacts that occasionally affected detection confidence, particularly for distant targets. Future optimizations targeting these bottlenecks could potentially reduce total response times to under 2 seconds while maintaining detection accuracy, proving critical for achieving real-time performance on the resource-constrained aerial platform.

\subsection{VR Performance Assessment}

The VR subsystem revealed significant performance constraints directly attributable to the Jetson Orin Nano's hardware limitations. Unlike other members of the NVIDIA Orin family, the Orin Nano lacks dedicated hardware encoding units (NVENC), forcing reliance on CPU-based software encoding solutions~\cite{nvidia_orin_nano_software_encode_2024,ridgerun_software_encoders_2023}. This architectural limitation created a fundamental bottleneck in the VR streaming pipeline that significantly impacted user experience quality.

The Jetson Orin Nano could only sustain 5 Mbps output through software-based H.264 encoding, well below the 60 Mbps required for smooth 4K streaming~\cite{VrARApplications}. This constraint resulted in visible compression artifacts and a severely compromised frame rate of 0.5 fps, falling dramatically short of the 30 fps target necessary for acceptable VR experiences~\cite{milvus_vr_frame_rate_2025}. The encoding bottleneck consumed approximately 99\% of CPU resources per core when using the ultrafast x264 preset, the only configuration capable of running in the Orin Nano. ~\cite{nvidia_orin_nano_software_encode_r36_2024}~\cite{nvidia_orin_nano_software_encode_r35_2023}.

Testing revealed that software encoding presets presented stark trade-offs between quality and performance. The ultrafast preset achieved the highest throughput at 0.5 fps for 4k streams, but with significant quality degradation, while medium and slow presets produced superior visual quality but dropped to 1 frame every 20 seconds. ~\cite{nvidia_orin_nano_software_encode_r35_2023}~\cite{nvidia_orin_nano_software_encode_r36_2024}. For the 360° video streaming application, the ultrafast preset was the only viable option, explaining the observed quality limitations and compression artifacts in the VR experience.

The software encoding process consumed substantial computational resources that competed with other system functions. CPU utilization reached 99\% during intensive encoding operations, leaving limited processing capacity for concurrent AI inference and network communication tasks~\cite{ridgerun_jetson_orin_nano_2024}~\cite{nvidia_orin_nano_software_encode_r36_2024}. This resource contention contributed to the overall system latency of 10.2 seconds, as the Jetson struggled to balance YOLOv11 processing, video encoding, and network transmission simultaneously.

Evaluation of different encoding approaches revealed that FFmpeg and GStreamer software encoders achieved similar performance limitations on the Orin Nano platform. Both frameworks could sustain real-time encoding only with the ultrafast preset, confirming that the bottleneck was hardware-imposed rather than software-specific~\cite{nvidia_orin_nano_software_encode_r36_2024}. Custom CUDA-based encoding solutions were considered but deemed impractical due to development complexity and uncertain performance gains~\cite{ridgerun_software_encoders_2023}.

Field testing confirmed that the 5 Mbps encoding limitation directly impacted VR usability, with operators reporting motion sickness and difficulty interpreting visual information due to compression artifacts and low frame rates~\cite{arborxr_vr_motion_sickness_2024}. Optimization attempts, including bitrate reduction, resolution scaling, and frame rate limiting, provided marginal improvements but could not overcome the fundamental hardware constraint. The average end-to-end latency of 10.2 seconds, while acceptable for non-critical monitoring, highlighted the need for hardware-accelerated encoding in future platform iterations~\cite{byteplus_vr_latency_2025}.

These findings demonstrate that while the Jetson Orin Nano provides excellent AI processing capabilities for edge computing applications, its lack of hardware video encoding significantly limits its suitability for high-quality VR streaming applications. Future implementations targeting immersive experiences should prioritize platforms with dedicated encoding hardware to achieve acceptable performance standards.

\section{Conclusion}\label{sec:conclusion}

This article has detailed a comprehensive UAV system for non-terrestrial networking, uniting immersive visualization, advanced analytics, and intelligent assistance to improve situational awareness. Leveraging 5G, the platform achieves low-latency object detection, panoramic imaging, and VR-based scene interpretation. LLMs enrich data with context, refining operator decision-making.

Experiments confirm the system’s utility in disaster relief, infrastructure inspection, and environmental monitoring. However, challenges remain in encoding and network efficiency. Future work will focus on pipeline optimization, hardware acceleration, and broader application support. The described methods are extensible to domains such as security and precision agriculture, highlighting UAVs’ versatility as intelligent NTN nodes.

\section{Future Work and Research Directions}\label{sec:future_work}

Future iterations should prioritize hardware-accelerated video encoding to overcome current VR streaming limitations. Integration of NVIDIA Jetson Orin NX variants with dedicated NVENC encoders could achieve the required 60 Mbps throughput for smooth 4K streaming. Advanced 5G mmWave connectivity could reduce the current 10.2-second latency to sub-second response times, enabling real-time immersive operations.

The platform's modular architecture enables expansion toward autonomous mission planning using LLM-driven natural language interfaces. Multi-UAV coordination systems leveraging distributed AI processing and shared contextual understanding could enable complex collaborative missions. Integration of edge-based quantized LLM models could reduce cloud dependency while maintaining analytical capabilities.

\section{ACKNOWLEDGMENTS}
The authors would like to acknowledge the support of Spanish projects ITACA (PDC2022-133888-I00), 6G-INTEGRATION-3 (TSI-063000-2021-127), and Fun4Date (PID2022-136684OB-C21) and the EU SNS project SEASON (grant 101096120) for the development of this work.

\bibliographystyle{IEEEtran}
\bibliography{references}

\end{document}